\documentclass{article}
\usepackage{amssymb,amsmath,hyperref}
\usepackage[cm]{fullpage}

\begin{document}

\begin{figure}
\begin{flushright}
TIFR/TH/11-42
\end{flushright}
\end{figure}

\title{Low energy supersymmetry from R-symmetries}
\author{Zheng Sun\\
        \normalsize\textit{Department of Theoretical Physics, Tata Institute of Fundamental Research,}\\
        \normalsize\textit{1 Homi Bhabha Road, Mumbai 400005, India}\\
        \normalsize\texttt{E-mail: zsun@theory.tifr.res.in}}
\date{}
\maketitle

\begin{abstract}
In a generic setting of Wess-Zumino models, we prove that the existence of a supersymmetric vacuum with a vanishing superpotential can be a consequence of a continuous or discrete R-symmetry when invariant fields are not less than fields transforming in the same way as the superpotential under the R-symmetry. The realization in string theory is discussed. We show that a rich landscape of low energy supersymmetric vacua can be found in the Type IIB flux compactification setup ready for the KKLT construction of de Sitter vacua in string theory.
\end{abstract}

\section{Introduction}

Supersymmetry (SUSY), along with its breaking and mediation mechanism, has been proposed for many years to solve several puzzles of the standard model, including the hierarchy problem and gauge coupling unification \cite{Martin:1997ns, Giudice:1998bp, Intriligator:2006dd, Intriligator:2007cp, Meade:2008wd, Dine:2009gy, Kitano:2010fa, Dine:2010cv}. In the gauge mediation scenario, the most interesting models have SUSY breaking at a scale not far from a TeV\@. And gravity mediation prefers intermediate scale SUSY breaking. One of the main tasks in fundamental physics is to understand different mass hierarchies, such as the one between the Planck scale and the scale of SUSY breaking. While building models from fundamental theories such as string theory, SUSY vacua with vanishing superpotentials are especially interesting. They lead to a zero vacuum energy in the supergravity (SUGRA) version of Wess-Zumino models, allowing other small effects to tune the cosmological constant to the observed value. One can construct dynamical SUSY breaking models from these vacua, and the dynamical scale statistically prefers the lowest value which is welcome to gauge mediation \cite{Banks:2003es, Dine:2004is}. Such vacua have been often observed as the result of discrete R-symmetries in flux compactification \cite{Dine:2004is, Dine:2004ct, DeWolfe:2004ns, Dine:2005yq, DeWolfe:2005gy, Dine:2005gz, Dine:2007er, Dine:2008jx}. Meanwhile, R-symmetries, especially discrete ones, account for various aspects of low energy phenomenology \cite{Dine:2010eb, Dine:2009swa, Lee:2010gv, Lee:2011dya, Asano:2011zt, Evans:2011mf}. The relation between R-symmetries and SUSY vacua for the special case with only R-charge $2$ and $0$ fields has been proved in previous literatures \cite{Dine:2005yq, Dine:2005gz, Dine:2007er, Dine:2008jx, Kappl:2010yu}. In this work, we provide a concrete proof for a sufficient condition to get SUSY vacua from R-symmetric superpotentials with fields of all possible R-charges. The proof works for both continuous and discrete R-symmetries. We show that this statement has important realization in Type IIB string theory flux compactification setup, with a rich landscape of SUSY vacua ready for the KKLT construction of SUSY breaking de Sitter vacua \cite{Kachru:2003aw}.

\section{SUSY vacua from R-symmetries}

Our setup is on a Wess-Zumino model whose coupling parameters have generic values. Such model serves as a low energy effective description of many theories. We categorize chiral fields into three types according to their behavior under the R-symmetry:
\begin{align}
R(X_i) &= R(W), \quad
i = 1, \dotsc , N_X;\\
R(Y_j) &= 0, \quad
j = 1, \dotsc , N_Y;\\
R(A_k) &\ne R(W) \text{ or } 0, \quad
k = 1, \dotsc , N_A.
\end{align}
If the R-symmetry is continuous then the superpotential $W$ has R-charge $2$. But the following proof also works for a discrete non-$\mathbb{Z}_2$ R-symmetry\footnote{A $\mathbb{Z}_2$ R-symmetry or R-parity does not change the superpotential. So there is no distinction between $X_i$'s and $Y_i$'s. Moreover, $\mathbb{Z}_2$ is not a true R-symmetry since the transformation of supercharges can be absorbed in a Lorentz rotation.}. So $R(W)$ should be understood as transforming in the same way as $W$, and R-charge $0$ should be understood as being invariant under the R-symmetry. Keeping every term transforming correctly, we can write down the most general form of the superpotential:
\begin{gather} \label{eq:2-01}
W = \sum_i X_i f_i(Y_j) + W_1,\\
\begin{split}
W_1 = &\sum_{\substack{i,j,k\\ R(A_k) = - R(W)}} \mu_{ijk} X_i X_j A_k
       + \sum_{\substack{i,j,k\\ R(A_j A_k) = 0}} \nu_{ijk} X_i A_j A_k\\
      &+ \sum_{\substack{i,j\\ R(A_i A_j) = R(W)}} \kappa_{ij} A_i A_j
       + \sum_{\substack{i,j,k\\ R(A_i A_j) = R(W)}} \lambda_{ijk} A_i A_j Y_k
       + \sum_{\substack{i,j,k\\ R(A_i A_j A_k) = R(W)}} \xi_{ijk} A_i A_j A_k\\
      &+ \text{non-renormalizable terms},
\end{split}
\end{gather}
where $-R(W)$ should be understood as transforming in the opposite way as $W$. All R-transformation requirements should be understood as satisfying the equality modulo $N$ in the case of a $\mathbb{Z}_N$ R-symmetry.

When searching for the vacuum, only scalar components of chiral fields enter the scalar potential. A scalar and its corresponding chiral field also transform in the same way under the R-symmetry. So we pick out scalars for fields in the following notation. Because of the R-symmetry, one can always find a quadratic factor $A_i A_j$ or $X_i X_j$ from each term of $W_1$ even if non-renormalizable terms are included\footnote{For a $\mathbb{Z}_{2 N}$ R-symmetry ($N \ge 2$), it is possible to have $R(W) = -R(W)$. In this case the first term of $W_1$ should be viewed as proportional to $X_i X_j X_k$. But the analysis works the same.}. So setting $X_i$'s and $A_i$'s to $0$, the form of $W_1$ ensures that all its first derivatives vanish.  The equations $f_i(Y_j) = 0$ can be solved for $N_Y \ge N_X$ unless $f_i$'s take some non-generic form. Then $W$ and all its first derivatives vanish. In summary, writing the superpotential in the form of \eqref{eq:2-01}, for $N_Y \ge N_X$ and generic $f_i$'s, one can find a SUSY vacuum with $W = 0$ by solving
\begin{equation}
f_i(Y_j) = 0, \quad
X_i = A_k = 0.
\end{equation}

Several remarks are to be addressed:
\begin{enumerate}
\item The Nelson-Seiberg theorem says that an R-symmetric superpotential is a necessary and sufficient condition for F-term SUSY breaking in generic models \cite{Nelson:1993nf}. What we present here is an exception since the (non-generic) R-charge condition $N_Y \ge N_X$ has to be satisfied.
\item It is known that if a superpotential with a continuous R-symmetry grants a SUSY vacuum, then the superpotential vanishes at the vacuum, although the R-symmetry can still be spontaneously broken by field expectation values. This can be generalized to approximate R-symmetry cases where the superpotential results to a small expectation value at the SUSY vacuum \cite{Kappl:2008ie}. It can also be viewed as a special (SUSY) case of the relation that the superpotential is bounded by the production of the R-axion and Goldstino decay constants \cite{Dine:2009sw}. However the relation does not tell whether such a SUSY vacuum exists or not. Our statement provides a sufficient condition for the existence of such a vacuum. R-symmetries are also generalized to discrete R-symmetries.
\item Although our model can include fields with any R-symmetry transformation properties, we have the SUSY vacuum which also preserves the R-symmetry since only $Y_i$'s, which are R-invariant, could acquire vacuum expectation values. By contrast, from a SUSY breaking vacuum in a model with a  continuous R-symmetry, fields with R-charges other than $0$ and $2$ are required for spontaneous R-symmetry breaking \cite{Shih:2007av, Carpenter:2008wi, Sun:2008va, Komargodski:2009jf, Curtin:2012yu}.
\item Similarly to the Nelson-Seiberg theorem, our result does not depend on whether the K\"ahler potential is R-symmetric or not. Only $W$ needs to be R-symmetric.
\item For SUGRA theories, $\partial_i W = W = 0$ implies $D_i W = 0$. So the vacuum preserves SUSY in both global SUSY and SUGRA cases. It also leads to a zero vacuum energy in SUGRA.
\end{enumerate}

\section{String theory realization}

Although the statement works for both continuous and discrete R-symmetries, it is the discrete version that has been found the most use in string phenomenology. Quantum gravity theories such as string theory do not allow global symmetries \cite{Hellerman:2010fv, Banks:2010zn}. But discrete symmetries are common. In the Type IIB flux compactification setup, 10-D string theory is compactified on a Calabi-Yau manifold to get a low energy 4-D $N = 1$ SUSY theory, and fluxes are turned on to generate the superpotential and stabilize moduli \cite{Gukov:1999ya, Grimm:2004uq, Grana:2005jc, Douglas:2006es}. The Calabi-Yau geometry can have many discrete symmetries under which the superpotential transforms. One can identify one of them as the R-symmetry, turn on only fluxes respecting the symmetry, and generate an R-symmetric superpotential \cite{DeWolfe:2004ns, DeWolfe:2005gy, Dine:2005gz}. The R-charge condition $N_Y \ge N_X$ is satisfied in many models. As a consequence, SUSY vacua with $W = 0$ can be found with generic choices of fluxes. By this means one can build a rich landscape of such vacua as the first step towards the KKLT construction of de Sitter vacua in string theory.

A simple and explicit example is the IIB theory on a $T^6 / \mathbb{Z}_2$ orientifold \cite{Kachru:2002he}. Looking for a vacuum with the complex moduli expectation values
\begin{equation} \label{eq:3-01}
\tau_{(0)} =
\begin{pmatrix}
i & 0 & 0\\
0 & \tau_{22} & \tau_{23}\\
0 & \tau_{32} & \tau_{33}
\end{pmatrix},
\end{equation}
there is a $\mathbb{Z}_4$ symmetry which rotates the first complex coordinate $z^1 = x^1 + i y^1$:
\begin{equation}
x^1 \to y^1, \quad
y^1 \to - x^1.
\end{equation}
The superpotential transforms like the holomorphic three-form $\Omega = dz^1 \wedge dz^2 \wedge dz^3$. So we have $W \to i W$ under the $\mathbb{Z}_4$. Using the notation in the previous section, there are four $X$-fields $\tau_{12}, \tau_{13}, \tau_{21}, \tau_{31}$ and five $Y$-fields $\tau_{22}, \tau_{23}, \tau_{32}, \tau_{33}, \phi$ where $\phi$ is the axion-dilation. We also have an $A$-field $\delta \tau_{11}$ defined as $\tau_{11} = i + \delta \tau_{11}$, with $\delta \tau_{11} \to - \delta \tau_{11}$ under the $\mathbb{Z}_4$. So there are more $Y$-fields than $X$-fields and our statement should apply. Indeed we can turn on invariant fluxes and have the superpotential
\begin{align}
W &= - (a_{12} - \phi c_{12}) (\operatorname{cof} \tau)_{12} - (b_{12} - \phi d_{12}) \tau_{12} + \{\text{terms with } 12 \leftrightarrow 13 , 21, 31\}\\
  &= \tau_{12} f_{12}(\tau_{22}, \tau_{23}, \tau_{32}, \tau_{33}, \phi) + \{\text{terms with } 12 \leftrightarrow 13 , 21, 31\}.
\end{align}
Setting $X$-fields to zero as suggested by \eqref{eq:3-01}, SUSY vacua with $W = 0$ can be found by solving $f$-functions for $Y$-fields. Since there are more $Y$-fields than $f$-functions, solutions do exist for generic choices of fluxes. The $A$-field $\tau_{11}$ does not enter the superpotential in this model. But $A$-fields may appear in more complicated models.

The $T^6$ manifold is not a general Calabi-Yau since it has a trivial holonomy instead of $SU(3)$. In fact the nine complex moduli $\tau_{ij}$ are redundant. Some of their combinations do not deform the manifold up to a change of K\"ahler moduli. This makes the field counting subtle. A Calabi-Yau with a full $SU(3)$ holonomy does not have such an issue. One example is the quintic surface in $\mathbb{CP}^4$ which is well explained in textbooks \cite{Green:1987mn}. Choosing the surface to be defined by the quintic equation
\begin{equation}
P = z_1^5 + z_2^5 + z_3^5 + z_4^5 + z_5^5 = 0
\end{equation}
where $z_1, \dotsc , z_5$ are coordinates of $\mathbb{CP}^4$, there are many symmetries which can be identified as R-symmetries. There are $101$ complex structure moduli corresponding to quintic monomial deformations of $P$. Their transformation property can be seen from these monomials. The holomorphic three-form $\Omega$ transforms like the monomial $z_1 z_2 z_3 z_4 z_5$, and so does the superpotential. If we choose the $\mathbb{Z}_5$ symmetry
\begin{equation}
z_1 \to e^{2 \pi i / 5} z_1
\end{equation}
and check the transformation properties of quintic monomials, there are $31$ $X$-fields, $41$ $Y$-fields including the axion-dilation and $30$ $A$-fields. Again we see $Y$-fields are more than $X$-fields, and generic choices of fluxes lead to SUSY vacua with $W = 0$.

The quintic in $\mathbb{CP}^4$ model can be generalized to the polynomial surface in the weighted projective space $\mathbb{WCP}^4$. A huge set of $\mathbb{WCP}^4$ spaces labeled by their weights of coordinates as well as choices of polynomials has been constructed as Calabi-Yau spaces \cite{Klemm:1992bx, Kreuzer:1992da, cy:url}. As a demonstrative example, consider in the space $\mathbb{WCP}^4_{1,1,1,6,9}$, the surface is defined by the polynomial equation
\begin{equation}
P = z_1^{18} + z_2^{18} + z_3^{18} + z_4^3 + z_5^2 = 0,
\end{equation}
and the $\mathbb{Z}_3$ R-symmetry is determined by
\begin{equation}
z_4 \to e^{2 \pi i / 3} z_4.
\end{equation}
Following the previous counting procedure, we find $91$ $X$-fields, $182$ $Y$-fields and no $A$-field. One can choose different R-symmetries and $N_Y \ge N_X$ is usually satisfied. A counterexample is the $\mathbb{Z}_{18}$ symmetry
\begin{equation}
z_4 \to e^{2 \pi i / 3} z_4, \quad
z_1 \to e^{\pi i / 9} z_1.
\end{equation}
There are $24$ $X$-fields, $17$ $Y$-fields and $232$ $A$-fields. So we see $N_Y < N_X$ in this (rarely found) case. Surveying the list of $\mathbb{WCP}^4$ spaces, it turns out that $N_Y \ge N_X$ is rather easily satisfied in most models with proper choices of R-symmetries. Meanwhile, much more than half of fluxes transform under the R-symmetry and have to be turned off in typical models \cite{Dine:2005gz}. This suggests that the SUSY vacua of our interest, while well-behaved, are not statistically favored. One has to seek non-perturbative stability or phenomenology reasons to select such vacua from the landscape \cite{Dine:2007er, Dine:2008jx, Dine:2009tv}.

In these examples, finding SUSY vacua with $W = 0$ has much importance: They contribute to ``the third branch'' of the landscape \cite{Dine:2005yq}, which is the only branch with unbroken supersymmetry at tree level. Vacuum statistics on this branch prefers a low scale for SUSY breaking when SUSY breaks dynamically \cite{Banks:2003es, Dine:2004is}. The equation $W = 0$ is actually the no-scale version of the SUSY equation $D_\rho W = 0$ where $\rho$ is a K\"ahler modulus. Since $\rho$ does not appear in $W$ at tree level, the number of SUSY equations is always one more than the number of fields. So it is hard to find a solution in general but here the R-symmetry helps. Following the steps of KKLT \cite{Kachru:2003aw}, K\"ahler moduli enter $W$ non-perturbatively making the vacuum to be anti de Sitter, then anti $D3$ branes lift up the vacuum to de Sitter and break SUSY\@. Other SUSY breaking effects can also be introduced (at low energy). In summary, R-symmetries play an essential role in building low energy SUSY models in string phenomenology studies.

\section*{Acknowledgement}

The author would like to thank Michael Dine, Sandip Trivedi and Satoshi Nawata for helpful discussions.

\end{document}